\documentclass[final,5p,times,twocolumns]{elsarticle}
\usepackage{moreverb,url}
\usepackage[colorlinks,bookmarksopen,bookmarksnumbered,citecolor=red,urlcolor=red]{hyperref}
\usepackage{graphics}
\usepackage{amsmath}
\usepackage{graphicx}
\usepackage{caption}
\usepackage{amssymb}
\usepackage{multirow}
\usepackage{tikz}
\usetikzlibrary{shapes.geometric, arrows,calc}

\usepackage[final, addedmarkup=bf]{changes}
\setaddedmarkup {\textcolor{violet}{#1}}
\newcommand\BibTeX{{\rmfamily B\kern-.05em \textsc{i\kern-.025em b}\kern-.08em
T\kern-.1667em\lower.7ex\hbox{E}\kern-.125emX}}

\usepackage{numcompress}
\journal{Journal of Materials Research and Technology}

\begin{document}

\begin{frontmatter}

\title{Multiphase Aluminum A356 Foam Formation Process Simulation Using Lattice Boltzmann Method}

\author[tehran]{Mojtaba Barzegari\corref{mycorrespondingauthor}}
\cortext[mycorrespondingauthor]{Corresponding author}
\ead{mbarzegary@alumni.ut.ac.ir}

\author[amirkabir]{Hossein Bayani}
\ead{hossein.bayani@aut.ac.ir}

\author[amirkabir]{Seyyed Mohammad Hosein Mirbagheri}
\ead{smhmirbagheri@aut.ac.ir}

\author[conc]{Hasan Shetabivash}
\ead{h\_sheta@encs.concordia.ca}

\address[tehran]{Department of Biomedical Engineering, Faculty of New Sciences and Technologies, University of Tehran, Tehran, Iran}
\address[amirkabir]{Department of Mining and Metallurgical Engineering, Amirkabir University of Technology, Tehran, Iran}
\address[conc]{Department of Mechanical and Industrial Engineering, Concordia University, Montreal, Quebec, Canada}

\begin{abstract}
Shan-Chen model is a numerical scheme to simulate multiphase fluid flows using Lattice Boltzmann approach. The original Shan-Chen model suffers from inability to  accurately predict behavior of air bubbles interacting in a non-aqueous fluid. In the present study, we extended the Shan-Chen model to take the effect of the attraction-repulsion barriers among bubbles in to account.  The proposed model corrects the interaction and coalescence criterion of the original Shan-Chen scheme in order to have a more accurate simulation of bubbles morphology in a metal foam. The model is based on forming a thin film (narrow channel) between merging bubbles during growth. Rupturing of the film occurs when an oscillation in velocity and pressure arises inside the channel followed by merging of the bubbles. Comparing numerical results obtained from proposed model with mettallorgraphy images for aluminum A356 demonstrated a good consistency in mean bubble size and bubbles distribution. 
\end{abstract}

\begin{keyword}
Metal Foam \sep  Aluminum A356 \sep  Form Grip \sep  Lattice Boltzmann Method \sep  Shan-Chen Model \sep  Multiphase Fluid Dynamics
\end{keyword}

\end{frontmatter}

\section{Introduction}
\label{int}
Demanding for advanced materials are increasing rapidly via new technologies. Closed cell metal foams gained a lot of interest as one of the major branches of advanced materials due to their unique physical and mechanical properties, including high specific strength and compressibility along with good energy absorption capability \cite{banhartjom2000,banhart2001,banhart1995,koernerbook}.

Despite the advantages, the employment of metal foams in industrial applications is limited due to the inhomogeneity of the structure which results in the deviation of the mechanical properties of the foams from what predicted by the scaling relations. This is mainly due to the morphological defects such as missing or wavy distortions of the cell walls and non-uniform shape and size of the cells which results in poor reproducibility of foam structures \cite{koernerpaper}. \deleted{The foams are quite stable in the absence of	 external factors but bubbles stability and energy balance in cell structures would be considered as  first challenges in production procedure. To overcome these difficulties, one should consider improving the governing mechanisms according to the understanding of basic physical phenomena occur during metal foam formation} 
In metals, unlike ionic liquids, the formation mechanisms of metal foam has not yet fully understood \cite{koernerpaper}.

Bubble stability is the primitive challenge in understanding the mechanism of metal foam formation. A variety of studies and researches have been performed by scientists in order to investigate and analyze the parameters affecting bubble stabilization \added{\cite{koernerbook, koernerpaper, Stanzick2002, 8-koerner, cavitation, RefA1, RefB1, RefB2, RefB3, RefB4, RefEmul}}.
Most of the investigations have focused on formation of single bubble in ionic liquid environment, especially water, with no impurity \added{\cite{cavitation, RefB1, RefB2, RefB3, RefB4, RefEmul}}. 
However, even in the purest condition, molten metal contains dozens of different impurities. Another important issue which has been neglected during various studies is the multi-bubble nature of metal foam formation process, which mostly appears in bubbles interactions. In addition, due to the presence of metallic bond in metal melt, there is no ionic or polar attraction and repulsion forces, which causes a different behavior of the liquid-gas interface in metal melts in comparison with aqueous solutions \added{\cite{koernerbook, koernerpaper, Stanzick2002, 8-koerner}}.

In order to have a computational study on metal foam formation process, a basic understanding of bubble stability conditions in the presence of particles is required. Therefore, a computational model based on known dynamics of bubbles and improving it using the computational-experimental approach has to be built, which is accomplished by adding some constraints that are focused on the boundary of each bubble according to available theories and verify the selected ones using an experiment \added{\cite{koernerpaper}}.

Most of the studies in the field of bubble dynamics investigated single bubble dynamics in aqueous solutions or water based liquids and only a few are conducted to study multi bubble dynamics and foam formation process.

Chine and Monno \cite{RefA1} developed an axial symmetry model to simulate the behavior of a single gas bubble expansion, embedded in a viscous fluid using Finite Element Method. Ghosh and Das \cite{RefA2} conducted  a numerical investigation of single bubble dynamics using Lattice Boltzmann Method. Their model contains a rising gas bubble inside a tube filled with liquid. They validated and verified different aspects of using Lattice Boltzmann Method in simulating trapped gas bubbles in liquids .

Chahine \cite{RefB1} studied the dynamics of clouds of bubbles via both analytical technique and numerical simulation using 3D Boundary Element Method (BEM). They also studied the behavior of bubbles in a non-uniform flow field and their response to flow. Chahine \cite{RefB2} investigated influence flow field to interaction of bubbles. Ida \cite{RefB3} conducted a mathematical modeling using a nonlinear multi-bubble model from pressure pulses perspective. Bermond et \replaced{al.}{al} \cite{RefB4} performed an outstanding computational investigation on bubble interactions and validated it using a novel experimental procedure. The dynamics of bubbles in that research was studied based on Rayleigh-Plesset equation. Kim et \replaced{al.}{al} presented an Immersed Boundary Method (IBM) to simulate and predict the 2D structure of a dry foam. This method allows one to study pressure equilibrium in non-conventional approach. In their model, a set of thin boundaries partition the gas into discrete cells or bubbles \cite{RefF1}.

The Lattice Boltzmann (LB) simulation has been used extensively to simulate the kinetic effects on bubbles \cite{Gupta2008}. Advantage of LB approach lies in the fact that there are no global systems of equations which have to be solved. Besides, boundaries in simulation domain do not have an effective impact on the computation time. These features are essential for foam formation simulation regarding the complex internal structure of foams.  Andrel et \replaced{al.}{al} \cite{RefC1} used Lattice Boltzmann to simulate flow in a simplified single phase model as an alternative to a liquid-gas two phase model to analyze bubble interaction in protein foams in order to determine bubble coalescence conditions. Leung et \replaced{al.}{all} \cite{RefD1} studied bubbles nucleation, growth, stability conditions and interaction in plastic foaming process. 

Beugre et \replaced{al.}{al} \cite{RefE2} developed a 3D Lattice Boltzmann code to simulate fluid flow in metal foam. Pressure drop was the criteria used to compare the obtained results with experimental measurements. Computer aided X-ray microtomography was used to produce the 3D geometry of metal foam imported in LB simulation. They improved the computed geometry of the metal foam later with more advanced techniques in order to obtain better results . 

K\"orner \cite{koernerpaper} conducted a thorough research using Lattice Boltzmann method to simulate the growth and formation of an aluminum metal foam using a single phase model. Diop et \replaced{al.}{al} simulated solidification process of metal foams using Lattice Boltzmann Method \cite{RefF5}. Their numerical model included aluminum melt with gassing agent heated to obtain metallic foam.

In this study we constructed a hybrid experimental-computational model to predict the structure of a two-dimensional closed-cell aluminum \deleted{metal} foam.  Most of the simulations conducted in this field are based on single phase models, while a two-phase computational model is adopted in the present study. To this end, a modified and improved Shan-Chen model is developed for multiphase simulations. Besides, a novel boundary condition is utilized to simulate oxide network particles' effect on the interaction of two bubbles. This method can be extended to account for a larger number of bubbles in order to simulate metallic foams. The proposed model is created based on an experimental procedure and actual samples' structures. Moreover, some additional phenomena such as random \deleted{bubble} nucleation, \deleted{bubble} growth\added{,} \deleted{and bubble} coalescence\added{, and aging of bubbles} are also developed \deleted{and included} in this model.

\section{Mathematical Modeling}
\label{commod}

It is assumed that there is no variation in liquid temperature during foaming process, and the fluid flow is considered to be incompressible. Consequently, conservation of mass and momentum in a single phase continuum can be written as follows \cite{JohnCFD}:

\begin{equation}
\nabla . \textbf{u} = 0
\end{equation}

\begin{equation}
\frac{\partial \textbf{u}}{\partial t} + \left(\nabla . \textbf{u} \right) \textbf{u} = -\frac{1}{\rho} \nabla p + \nu \nabla^2 \textbf{u} + g
\end{equation}
where $ t $,  $ u $, $ p $, $ \rho $, $ \nu $, $ g $ are  time, velocity,   pressure,   density,   kinematic viscosity and  gravity, respectively.

In addition to conservation equations, gas pressure ($P_i$) in bubble ($ i $) could be expressed by ideal gas equation \cite{cavitation}:

\begin{equation}
{p_i} = \frac{{{n_i}.R.T}}{{{V_i}}}
\label{equ3}
\end{equation}
where $ R $ is gas constant,  $ n_i $  gas mole, $ T $  temperature and $ V_i $  is volume of bubble $ i $. Gas and liquid were coupled at the interface by momentum balance and controlled by $ \Gamma $ parameter. For having a stable bubble, velocity of fluid and gas must be equal at the interface.

\begin{equation}
{v_G}(x) = {v_F}(x) \qquad \forall \,x \in \Gamma
\label{equ4}
\end{equation}

In order to estimate the pressure during the growth or expansion of the bubble, the kinetics of expansion of a single bubble of radius $R$ in an incompressible fluid with viscosity of $\rho$ and pressure $p_0$ is considered as basic calculations and due to the symmetry of the problem, the NSE equation can be reduced to Rayleigh equation\cite{cavitation} in spherical coordinates. Thus the bubble growth with radius $R$ can be calculated from Eq. \ref{equ7}.

\begin{equation}
\rho R\mathop R\limits^{..} + \frac{3}{2}\rho \mathop {{R^2}}\limits^. \quad + 4\rho \nu \frac{{\mathop R\limits^. }}{R} \quad+ \frac{{2\sigma }}{R} = {p_i} - {p_0}
\label{equ7}
\end{equation}

where $ \mathop R\limits^{.}  $ is $ \frac{\partial R}{\partial t} $ and $ \mathop R\limits^{..}  $ is $ \frac{\partial^2R}{\partial t^2} $. Each of the four terms of the Eq. \ref{equ7} from left to right represents the excess bubble pressure, capillary pressure, viscous pressure and inertia pressure where $p_i$ denotes the bubble pressure and $\sigma$ is the surface tension. The Rayleigh equation expresses equilibrium between inertia, viscous and capillary forces, which prevent bubble expansion. \replaced{However,}{But} in practice\deleted{, which of} these forces will play a major role \added{that} is unclear and entirely depends on foam material and process parameters \cite{koernerbook}.

In this study it is assumed that blowing agent with \replaced{specified}{a specific} weight percent was added to molten metal, uniformly distributed and dissolved by stirring. Then \replaced{micro}{the} bubbles \deleted{that are very small and not visible to the naked eye (bubble nuclei)} were abruptly produced by dissolution reaction of blowing agent \replaced{in throughout the domain}{around the melt}. 

Nucleation occurs in random points \replaced{in the}{of} domain. After nucleation, bubble growth will begin. For growth modeling in each time step, gas will be added by virtual blowing agent (proportional to blowing agent weight and gas production rate) in each bubble, and then the bubble volume will be increased \replaced{due to the}{because of} pressure balance. Gas blowing or growth will continue until all virtual blowing agent gases are added to the domain. Other phenomena\added{,} such \added{as} drainage and wall rupture are considered in growth process by \deleted{equilibrium} bubble \added{equilibrium} equations. \replaced{Condition of}{Because of} wall rupture \deleted{additional condition is added along the way. This condition} could be \added{added as} \deleted{either} an experimental \deleted{one} or a \deleted{computational and} mathematical \replaced{criterion}{condition}. One would consider this condition experimentally. For example, if the cell wall thickness falls below a critical thickness (which is a material characteristic), cell wall rupture will occur. This critical number for aluminum melt is reported as $50 \mu m $, determined by X-ray radiography \cite{Stanzick2002}. In \replaced{this}{present} study, a computational \replaced{criterion}{condition} \deleted{is utilized, discussed in \replaced{page}{section} \pageref{interaction}, which is the determination of cell wall rupture using the} \added{will be obtained from} second derivative of pressure ($\triangledown^2 p$).

Hydrodynamics, gas release and diffusion are necessary for the foaming stage. The blowing gas solubility in the fluid is finite. The gas concentration in the fluid can be calculated by the diffusion equation:
\begin{equation}
{\partial _t}c + {\upsilon _x}{\partial _x}c - {\partial _x}(D{\partial _x}c)) = Q
\label{equ8}
\end{equation}
where $c = c(x, t)$ is concentration field \added{of dissolved gas}, $D$ gas diffusion constant and $Q = Q(x, t)$  is source term. The source term ($  Q$) describes the blowing agent decomposition. There are two key points which have to be considered. One is gas solubility and the other is gas diffusion distance. In this study the \replaced{concentration}{amount} of dissolved hydrogen in molten and solid aluminum \added{(C)} is calculated from Eq. \ref{equ9} and Eq. \ref{equ10} respectively \cite{koernerbook}:
\begin{equation}
C = 5.84\frac{{c{m^3}}}{g}.\exp( - \frac{{6357K}}{T})\sqrt {\frac{p}{{bar}}} \qquad  \small\text{For melt }
\label{equ9}
\end{equation}
\begin{equation}
C = 0.25\frac{{c{m^3}}}{g}.\exp ( - \frac{{5941K}}{T})\sqrt {\frac{p}{{bar}}} \qquad \small\text{For solid}
\label{equ10}
\end{equation}

Diffusion distance $\delta^{dif}$  is calculated from Eq. \ref{equ11}  where $D$ denotes the diffusion coefficient and  $t$ the characteristic time \cite{koernerbook}:
\begin{equation}
{\delta ^{dif}} = \sqrt {4Dt}
\label{equ11}
\end{equation}
The diffusion length is a measure \replaced{for}{of} the \added{region of influence of a} blowing agent particle\deleted{ influence region}. For example\added{,} if $\delta^{dif}$ is significantly larger than the mean particle distance, then a strong mutual influence has to be expected.
The diffusion length is a measure for the region of influence of a blowing agent
particle.

In this study, diffusion coefficient of hydrogen in aluminum is calculated from Eq.  \ref{equ12} \cite{koernerbook}:
\begin{equation}
D = {D_0}.\exp( - \frac{H}{{RT}}) = \left\{ {\begin{array}{*{20}{c}}
{3.8.{{10}^{ - 6}}\frac{{{m^2}}}{s}.\exp( - \frac{{19.26kJ/mol}}{{RT}})} \qquad \\
{1.1.{{10}^{ - 5}}\frac{{{m^2}}}{s}.\exp( - \frac{{40.95kJ/mol}}{{RT}})} \qquad 
\end{array}} \right.
\label{equ12}
\end{equation} 
where $D_0$ is a constant, $R$ is the gas constant, $T$ is the temperature and $H$ the activation enthalpy. From Eq. \ref{equ12}, the diffusion coefficient at 700 $ ^\circ C $ is $D_{700^\circ C} = 3.51\times 10^{-7} \frac{m^2}{s}$ 
  and diffusion length of hydrogen is $ 374\mu m $. This length is small compared to the overall dimension of casting. Thus, gas concentration uniformity cannot be expected. Consequently\added{,} inhomogeneous distribution of blowing agent in melt will lead to \replaced{non-uniform porosity}{inhomogeneous porosities}. Experimental observations have confirmed this theory\deleted{. Because to inhomogeneous distribution of blowing agent, foamed areas can be adjoined directly to unfoamed areas} \cite{koernerbook}.

\subsection{Numerical Model}
\label{numerical}

\subsubsection{Lattice Boltzmann approach}
\label{lbm}	

The \replaced{LB}{lattice Boltzmann} method \deleted{(LBM)}	 has shown to be suitable for foam formation problems \cite{succibook}. Random micro bubbles in a virtual medium nucleate and interact within a set of rules. If correct physics is applied in the simulation, spontaneous hydrodynamic behavior can be expected. It can be said that LB\deleted{M} method is a mesoscopic approach that is between macroscopic CFD approaches and microscopic molecular dynamics. Many multiphase models  exist that use the LB\deleted{M} \added{method}, such as Immiscible Lattice Boltzmann (ILBM) \cite{RefGrunau1}  \replaced{Shan-Chen}{ShanChen} Model \cite{MS30,MS31}, Free energy model \cite{MS32,MS33}, Chromodynamic model \cite{MS29} and HSD model \cite{MS34}.

\replaced{LB}{Lattice Boltzmann} method models fluid dynamics by evaluating particle distribution function $ f(x,v,t) $  at each lattice point, where $ f $ is the probability of finding a moving fluid particle with velocity $ v $ at point $  x$ and time $  t$. By knowing $  f$, one could get the values of density and momentum. The distribution function used in LB\deleted{M} method, $ f_i $, is a discretized form of the main continues function. Discretization means dividing the space into a finite number of lattices in order to present different parameters on these points, e.g. velocities could be evaluated by displacement vectors as ($\delta t .\replaced{\textbf{e}}{e} _i$) where ($ \delta t $) is time step and $  i$  is displacement direction.

At each lattice point, different sets of distribution function would be defined. Two mostly used functions are $  f$ and $  h$. The $ f $ function models mass and momentum transports and the $  h$ function perform energy transport modeling. The macroscopic parameters are given by aggregating these distribution functions \cite{RefShiyi1}:

\begin{equation}
\label{eq:vel_dens}
 \rho=\sum_i f_i , \quad \rho \textbf{u}= \sum_i \textbf{e}_i f_i, \quad E=\sum_i h_i
\end{equation}
where $ \rho $ is the density, $ \replaced{\textbf{u}}{u} $ the macroscopic velocity and $ E $ the energy density. As we have neglected thermal perturbation and solidification of liquid phase in present study, the energy transportations and its related functions will not be discussed further here.

The displacement of distributions \replaced{is}{are}  summarized by the equations of motion \cite{RefHe2}:

\begin{equation}
\label{eq:EOM}
f_i \left(\textbf{x}+e_i,t+\Delta t \right) - f_i \left(\textbf{x} , t \right) = \frac{\Delta t}{\tau_f}\left( f_i^{eq} \left(\textbf{x},t\right) \right) + F_i
\end{equation}
where $ f_i(\replaced{\textbf{x}}{x},t) $ is the density distribution function in i direction. In order to model external forces such as gravity, one would use  \cite{RefHe2}:

\begin{equation}
F_i=W_i \rho \left[ \frac{ \left( \textbf{e}_i - \textbf{u} \right)}{c_s^2}+ \frac{ \left( \textbf{e}_i \cdot \textbf{u} \right)}{c_s^2} \right]  \cdot \textbf{g}
\end{equation}

$f^{eq}_i(\replaced{\textbf{x}}{x},t)$ is equilibrium distribution function  \cite{RefHe2}:

\begin{equation}
f_i^{eq} \left( \textbf{x}, t \right) = w_i \rho \left[ 1+ \frac{ \left( \textbf{e}_\textbf{i} \cdot \textbf{u}  \right)}{c_s^2} + \frac{ \left( \textbf{e}_\textbf{i} \cdot \textbf{u}  \right)^2}{2c_s^4} - \frac{\textbf{u}^2}{2c_s^2}  \right]
\end{equation}

For a two dimensional D2Q9 model (2D with 9 velocity directions), the velocity direction $ \replaced{\textbf{e}}{e}_i $  and the weight $ \omega_i $ are given by  \cite{RefHe2}:

\begin{equation}
\textbf{e}_i = \left\{   \begin{array}{l l}   \left( 0, 0 \right), & \quad i=0\\   \left( \pm c, 0 \right), \left(0, \pm c \right) & \quad i=1, \cdots, 4\\ \left(\pm c, \pm c \right) & \quad i=5, \cdots, 8 \\  \end{array} \right.
\end{equation}

\begin{equation}
w_i = \left\{   \begin{array}{l l}   4/9, & \quad i=0\\  1/9 & \quad i=1, \cdots, 4\\ 1/36 & \quad i=5, \cdots, 8 \\  \end{array} \right.
\end{equation}

The viscosity $ \nu $  is given by:

\begin{equation}
\nu=c_s^2 \Delta t \left(\tau_f - 0.5 \right)
\end{equation}
where $ \tau_f $  is the relaxation time for velocity field in dimensionless form.

Equation of motion (Eq. \ref{eq:EOM}) is solved by LBM in two steps, as noted earlier, known as collision and streaming  \cite{RefHe2}:

\paragraph*{Collision:}

\begin{equation}
 f_i^{out} \left( \textbf{x}, t \right) = f_i^{in} \left( \textbf{x}, t \right) + \frac{\Delta t}{\tau_f} \left( f_i^{eq} \left( \textbf{x}, t \right) - f_i^{in} \left( \textbf{x}, t \right) \right) + F_i
\end{equation}

\paragraph*{Streaming:}

\begin{equation}
f_i^{in} \left( \textbf{x} + e_i , t+\Delta t \right) = f_i^{out} \left( \textbf{x}, t \right)
\end{equation}
where $ f^{out}_i $  and $ f^{in}_i $ are outgoing (after collision) and incoming (before collision) distribution functions respectively.

\subsubsection{Shan-Chen model}

Shan-Chen model is based on incorporating long-range attractive forces ($ \textbf{F} $) between distribution functions. In the original Shan-Chen model, the interacting force is approximated using the following equation \cite{MS30,MS31}:
\begin{equation}
F \left(x  \right) \cong -\frac{d^2b}{D} \psi \left(x \right) g \bigtriangledown  \psi \left(x \right) 
\label{shanchen1}
\end{equation}
Where $ b $ is the number of nearest sites with equal distance,  $ D $ is the dimension of the space and $ g $ is \replaced{proportional}{proportinal} to the interaction strength. The function $ \psi $ is so-called pseudopotential and is a function of time and location. Other adjacent sites (next nearest) can be considered in the Eq. \ref{shanchen1} which leads to a more general form of the above equation \cite{MS35}:
\begin{equation}
F \left(x  \right) = -c_0 \psi \left(x \right) g \bigtriangledown  \psi \left(x \right) 
\label{shanchen2}
\end{equation}

In the Shan-Chen model the force at a given lattice point depends on all local \replaced{neighbor's}{neighbours}	 characteristics. So the following can be written:
\begin{equation}
F_\alpha \left(x  \right) = -G \psi \left(x \right) \Sigma_i w_i  \psi \left(x+e_i \right)e_i 
\label{shanchen3}
\end{equation}
where coefficient $ G $ controls the strength of the attraction. The function  $ \psi $ is $ \psi=\psi \left( x \right) $, where $ \rho $  depends on time and location. $c_i$ and $w_i$ are respectively lattice velocity vector and its weight in selected lattice model. Therefore, force is introduced to account for attraction. The contributions of state and surface tension to the equation can be observed through Taylor expansion. Taylor expansion of the force can be written as \cite{PhDSbragaglia2007}:
\begin{equation}
F_\alpha \left((x)  \right) = -G \psi \left(x \right) \left(\frac{1}{3}\partial_\alpha\psi + \frac{1}{18}\partial_\alpha \Delta\psi \right) + O\left(\partial^5 \right)
\label{tylor}
\end{equation}

The following formulation is derived algebraically \cite{PhDSbragaglia2007}:

\begin{align}
F_\alpha \left((x)  \right) &= -G \psi \left(x \right) \left(\frac{1}{3}\partial_\alpha\psi + \frac{1}{18}\partial_\alpha \Delta\psi \right) \nonumber \\
&=-G \left(\frac{1}{6}\partial_\alpha\psi^2 + \frac{1}{18} \left( \partial_\alpha \left( \psi\Delta\psi \right) - \Delta\psi\partial_\alpha\psi \right) \right) \nonumber \\
&=-G \left(\frac{1}{6}\partial_\alpha\psi^2 + \frac{1}{18} \left( \partial_\alpha \left( \psi\Delta\psi \right) +\frac{1}{2}\partial_\alpha \vert \bigtriangledown \psi \vert^2  \right. \right. \nonumber \\
&\quad - \partial_\beta\partial_\alpha\psi \partial_\beta \psi \bigg) \bigg) 
\end{align}

The force influence can be included to in the momentum-flux tensor  \cite{PhDSbragaglia2007,MS35}:
\begin{equation}
\partial_\beta P_{\alpha \beta}=-F_\alpha + \partial_\alpha p
\end{equation}

The equation of state for the LBE is $ p=c^2_s \rho $ so:
\begin{equation}
\partial_\beta P_{\alpha \beta}=-F_\alpha + \partial_\alpha \left(c^2_s \rho \right) 
\end{equation}

Thus, the flux tensor $ P_{\alpha \beta} $ is modified as follows:
\begin{align}
P_{\alpha \beta}=\left( c^2_s\rho +\frac{G}{6}\psi^2 + \frac{G}{36} \vert \bigtriangledown\psi\vert^2 + \frac{G}{18}\psi\bigtriangleup\psi \right)\delta_{\alpha \beta} \nonumber \\ - \frac{G}{18}\partial_\alpha\psi\partial_\beta\psi
\end{align}

By analogy with classical mechanics, the potential of the force can be introduced as:
\begin{equation}
U=\frac{G}{6}\psi^2 + \frac{G}{36} \vert \bigtriangledown\psi\vert^2 + 	\frac{G}{18}\psi\bigtriangleup\psi
\label{UPotansial}
\end{equation}

Since the gradient term\added{s} in  Eq. \ref{UPotansial}  are \added{in} small compared to the leading terms (the characteristic length of the interface is longer than the lattice spacing, as in all diffuse-interface methods), the Eq. \ref{UPotansial} can be approximated:
\begin{equation}
p=\rho c^2_s + \frac{G}{6}\psi^2
\label{Pequ}
\end{equation}

By a suitable choice of the pseudo-potential $ \psi \left(x \right) $, this equation can describe the separation of phases. One simple and usual choice can be $ \psi = \rho $. By using this pseudo-potential function, the momentum flux tensor resembles diffuse interface method. However, the choice of $ \psi = \rho \left( x, t \right) $ is not the best in terms of stability. When $ \psi $ equals $\rho $, it becomes larger for larger $ \rho $. Thus, the attractive potential contains a malfunctioning loop: the larger density  $\rho $ leads to a larger $ \psi $  which causes larger gradients and instabilities. $ \psi = \rho $ is good for small gas-liquid density ratios.

\begin{figure}[htbp]
\centerline{\includegraphics[width=2in]{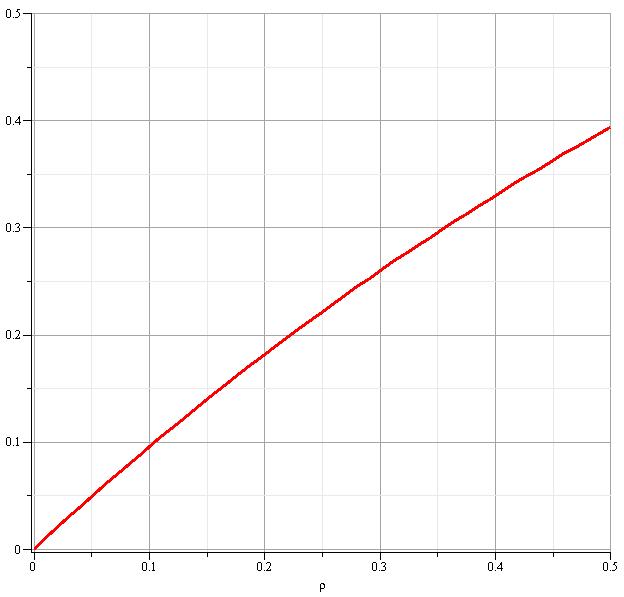}}
\caption{Pseudopotential function, zoomed to show its linearity for small densities}
\label{fig:psi-z}
\end{figure}

\begin{figure}[htbp]
\centerline{\includegraphics[width=2in]{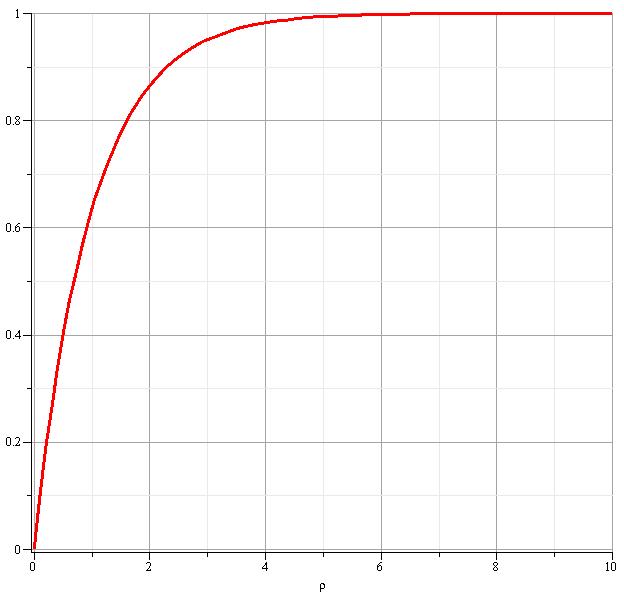}}
\caption{Pseudopotential function, which yields 1 as approaches infinity (large densities)}
\label{fig:psi}
\end{figure}

In the case of \deleted{two-phase system of} aluminum liquid and hydrogen gas \added{(two-phase system)}, \deleted{the} density ratio is considerably higher. Therefore\added{,} to handle the pseudopotential function $ \psi $ for larger $\rho$ while preserving its ratio for smaller $\rho$, the following choice of the pseudopotential is used:

\begin{equation}
\psi \left( \rho \right) = 1- exp\left( -\rho \right)
\label{pseudopotential}
\end{equation}
Which is for small $ \rho $ equals to $ \psi \left(x \right) = \rho $ (Fig.\ref{fig:psi-z}) and for large densities, $ \psi \left(x \right) = 1 $ (Fig.\ref{fig:psi}). This choice of the pseudopotential allows separation of gas and liquid in larger density ratios (if not more than 60-70) \cite{RefInamuro1}.

The critical value\added{,} \deleted{obtained} when \deleted{the} separation occurs\added{,} can be calculated from the thermodynamic theory by these two equations \cite{RefHe1}:

\begin{equation}
\frac{dp_0}{d\rho}=0
\label{dr}
\end{equation}
\begin{equation}
\frac{d^2 p_0}{d\rho^2}=0
\label{d2r}
\end{equation}
By substituting Eq. \ref{pseudopotential} in Eq. \ref{Pequ} and for D2Q9 lattice ($ c_s = \frac{1}{\sqrt{3}} $):
\begin{equation}
p_0=\frac{\rho}{3} + \frac{G}{6} \left(1- exp\left( -\rho \right)\right)
\label{D2Q9equ}
\end{equation}
And from Eq. \ref{dr} and Eq. \ref{d2r} one would get:
\begin{align}
&\frac{1}{3} + \frac{G_{critical}}{3} exp\left( -\rho_{critical} \right) \left(1- exp \left( -\rho_{critical} \right)\right)=0 \nonumber \\
&-\frac{G_{critical}}{3} exp\left(-\rho_{critical}\right)\left(1- exp \left( -\rho_{critical} \right)\right)\nonumber \\
&\qquad+\frac{G_{critical}}{3} exp\left(-2\rho_{critical}\right)=0
\end{align}

Solving \replaced{these}{this system of} equations \replaced{lead to}{to yield} \deleted{$ G_{critical} $ and $ \rho_{critical} $ will result to} $ G_{critical}=-4 $ and $ \rho_{critical}=ln2 $. This means \added{that} if the system is initialized with the liquid density more than ln 2 and the gas density less than ln 2 in simulations with $ G\leq-4 $, the result is stable and separation will occur.
\subsubsection{Bubble nucleation and growth}
Number of bubble nucleation sites depends on the initial amount and size of the blowing agents. This number which is based on experimental results, is initially inserted into the main procedure. A time random subroutine is used to determine nucleation site positions in a 2D domain. These locations called “domain gas points” are in fact virtual blowing agents which in lattice domain are defined as gas nuclei and hydrogen density resulting from decomposition of the blowing agents, is accordingly calculated for these points. Other lattice points are set as liquid and give aluminum density. After this step, all numbers  and parameter are changed to dimensionless parameters by open source OpenLB code \cite{openlb}.
Hydrogen gas release rate is calculated in each time step and added to each lattice point by pressure increment.
\begin{equation}
\frac{dp}{dt}= A \times \frac{\frac{dn}{dt}}{N}
\end{equation}
where $ \frac{dp}{dt} $  is pressure increase rate for each lattice point that refers to gas, $ A $ is a constant \added{which depends on gas behavior (in case of ideal gas $ A=\frac{RT}{V}$ where $R$ is gas constant, $T$ temperature and $V$ is volume)}, $ \frac{dn}{dt} $   is gas release rate in  $ \frac{mole}{sec} $ and $ N $ is population of lattice gas points. Thus, because of pressure increase, pressure expansion is defined by multiphase code that has been developed in this work.

\subsubsection{Program algorithm}

\begin{figure*}[htb]
\centering
\scriptsize
\begin{tabular}{*{15}{c}}

\tikzstyle{startstop} = [ellipse, rounded corners, minimum width=2cm, minimum height=1cm,text centered, draw=black, fill=gray!30]
\tikzstyle{io} = [trapezium, trapezium left angle=70, trapezium right angle=110, minimum width=2cm, minimum height=1cm, text centered, draw=black, fill=blue!30,text width=4.5em]

\tikzstyle{process} = [rectangle, rounded corners, minimum width=3cm, minimum height=1cm, text centered, draw=black, fill=green!30]
\tikzstyle{process2} = [rectangle, rounded corners, minimum width=3cm, minimum height=1cm, text centered, draw=black, fill=green!30, text width=10em]
\tikzstyle{process4} = [rectangle, rounded corners, minimum width=3cm, minimum height=1cm, text centered, draw=black, fill=red!30, text width=10em]
\tikzstyle{process3} = [rectangle, rounded corners, minimum width=3cm, minimum height=1cm, text centered, draw=black, fill=red!30]
\tikzstyle{decision} = [diamond, minimum width=3cm, minimum height=1cm, text centered, draw=black, fill=yellow!30,text width=4.5em, inner sep=0pt]
\tikzstyle{arrow} = [thick,->,>=stealth]
\tikzstyle{coord}=[coordinate, node distance=6mm and 25mm]

\begin{tikzpicture}[node distance=2cm]

\node (start) [startstop] {\text{Start}};
\node (f1) [io,below of=start, yshift=0.5cm] {Input DATA};
\node (f2) [process,below of=f1, yshift=0.5cm] {Geometry define};
\node (f3) [process3,below of=f2, yshift=0.5cm] {Lattice define};
\node (f4) [process,below of=f3, yshift=0.5cm] {Random Nucleation};
\node (f5) [process,below of=f4, yshift=0.5cm] {Define Material Properties};
\node (f6) [process3,below of=f5, yshift=0.5cm] {Define Boundary Condition};
\node (f7) [process,below of=f6, yshift=0.5cm] {Bubble growth};
\node (c7) [coord, right of= f7, xshift=2cm]   {};
\node (f8) [process3,right of=f1, xshift=3cm] {Collide step};
\node (f9) [process3,below of=f8, yshift=0.5cm] {Stream step};
\node (f13) [decision,below of=f9, yshift=0] {Coarsening condition};
\node (f10) [process2,below of=f13, yshift=0cm, xshift=0cm] {Modified Shan-Chen force calculation};
\node (f14) [process4,below of=f9, yshift=0cm, xshift=4cm] {Shan-Chen Force calculation};
\node (f11) [process3,below of=f10, yshift=0.5cm] {Domain coupling};
\node (f12) [process3,below of=f11, yshift=0.5cm] {Plot Results};
\node (f16) [process3,below of=f14, yshift=0.5cm] {Termination condition = True};
\node (f17) [decision,below of=f12, yshift=0] {Termination condition};
\node (c17) [coord, left of= f17, xshift=-1.8cm]   {};
\node (f15) [startstop,right of=f17, xshift=2cm] {Finish};

\draw [arrow] (start) -- (f1);
\draw [arrow] (f1) -- (f2);
\draw [arrow] (f2) -- (f3);
\draw [arrow] (f3) -- (f4);
\draw [arrow] (f4) -- (f5);
\draw [arrow] (f5) -- (f6);
\draw [arrow] (f6) -- (f7);
\draw [arrow] (f7.east) -- (c7) |- (f8);
\draw [arrow] (f8) -- (f9);
\draw [arrow] (f9) -- (f13);
\draw [arrow] (f13.east) --node[anchor=south, near start] {Yes} (f14);
\draw [arrow] (f13) --node[anchor=east, near start] {No} (f10);
\draw [arrow] (f14) -- (f16);
\draw [arrow] (f16) |- (f11);
\draw [arrow] (f10) -- (f11);
\draw [arrow] (f11) -- (f12);
\draw [arrow] (f12) -- (f17);
\draw [arrow] (f17.east) --node[anchor=south, near start] {True} (f15);
\draw [arrow] (f17.west) -- node[anchor=south, near start] {No} (c17)|-(f8);

\end{tikzpicture}

  \end{tabular}
\caption{Main Program Algorithm (Green: User Codes, Red: Open Source Codes)}
\label{fig:Mainflowchart}
\end{figure*}
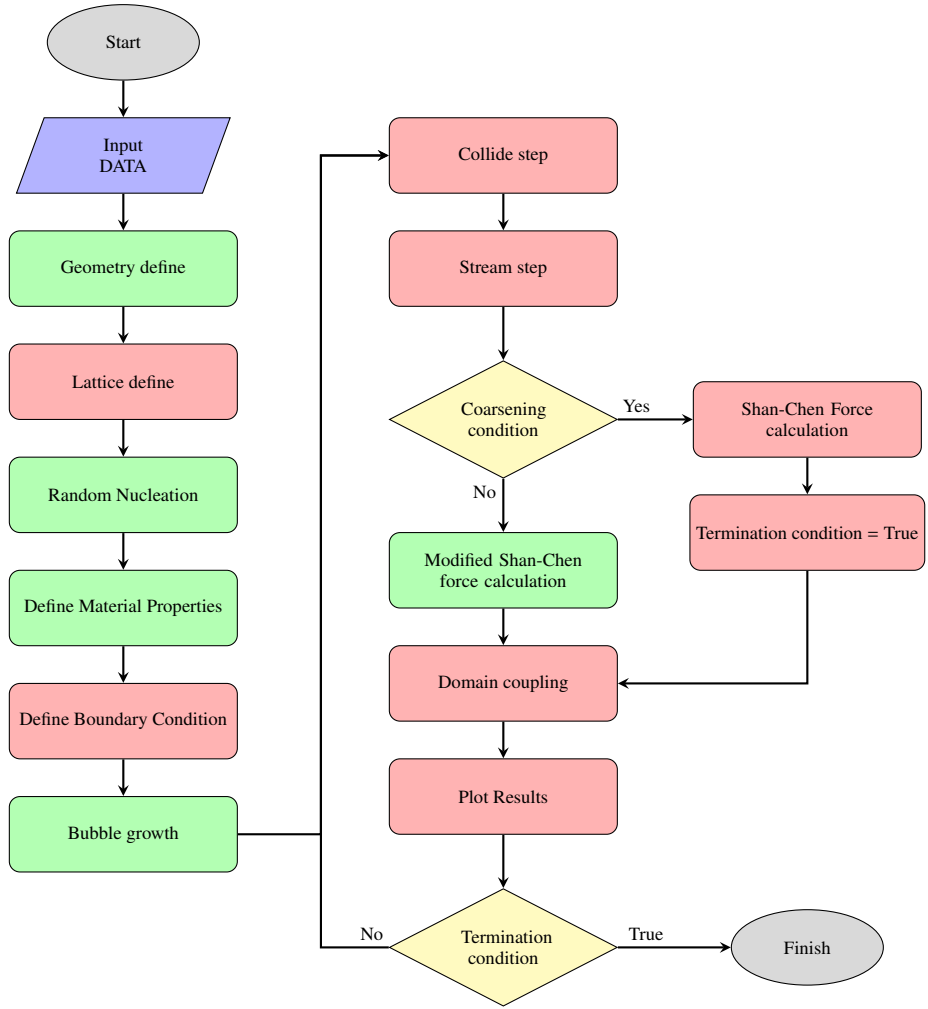

The flowchart of the main program algorithm is shown in Fig. \ref{fig:Mainflowchart}. \replaced{Present}{The} code \replaced{has been}{is} developed base on OpenLB open source code. The Shan-Chen model \replaced{was}{is} incorporated and  some modifications \replaced{were added}{are made} to the core structure of this algorithm. \deleted{This modification is described in \replaced{page}{section} \pageref{interaction}}. In Fig. \ref{fig:Mainflowchart}, the red boxes are the codes developed by open source community and the green boxes demonstrate the developed or modified \deleted{portions, performed in current} \added{parts, which where achieved in this} study.

\subsubsection{Bubbles interaction and modification of Shan-Chen model}
\label{interaction}

\begin{figure}[htbp]
\centerline{\includegraphics[width=3in]{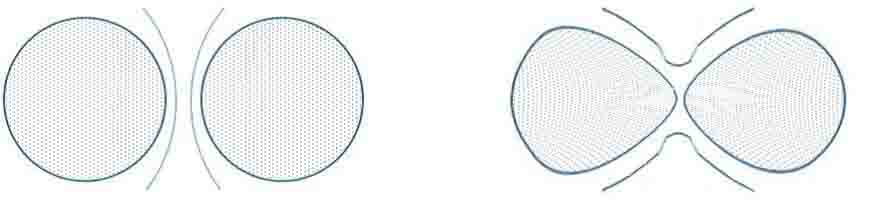}}
\caption{\replaced{Interaction}{Coalescence} of two bubbles in pure liquid \added{(interest to merging)}}
\label{fig:2}
\end{figure}

\begin{figure}[htbp]
\centerline{\includegraphics[width=3in]{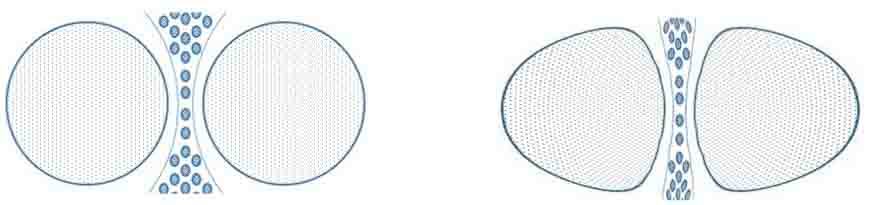}}
\caption{\replaced{Interaction of two bubbles in melt liquid including solid particles (tolerate to merging)}{Barrier effect of oxide network}}
\label{fig:3}
\end{figure}

Interaction of bubbles in pure liquid\deleted{s}\added{,} without suspended solid\added{,} modeling techniques is completely different from liquids containing floating particles (e.g. SiC particles in molten aluminum). \deleted{Naturally} \replaced{I}{i}n case of two moving bubbles in pure \added{aqueous} liquid\deleted{s}, if they \replaced{are}{come} close\added{d} to each other, \deleted{the most} common behavior is \deleted{the} increment in their surface curvature\added{, that leads to} \deleted{as they contact each other and start} merging \added{and coalescence phenomena} \replaced{from}{at} the contact tip (See Fig. \ref{fig:2}). \replaced{However, as shown in Fig. \ref{fig:3} for}{But in} molten metals, \replaced{due to existence of a lot of solid particles (impurities and inclusions), interaction between bubbles have a different behavior.}{as it is in the category on non-pure liquids, the situation is different.} Experimental observations \replaced{during aluminum}{of} foam\deleted{s} \replaced{production}{produced} \deleted{by powder compacts} show a particle\replaced{s}{-like} network \added{see Fig. \ref{fig:3}}  \deleted{that was formed} between bubbles, which \added{is often} called oxide network \cite{8-koerner}. This network \deleted{consists of particles located at the bubbles'}\added{at} interface\deleted{. They} act as a mechanical barrier and prevent further cell wall thinning. \replaced{Therefore,}{Thus} \added{main mechanism of foam stabilization} \deleted{particle confinement} between bubbles \added{is due to  particle confinement} \deleted{in metal foaming process can be regarded as the basic mechanism of stabilization }(See Fig. \ref{fig:3}).

\begin{figure*}
\centering
\scriptsize
\begin{tabular}{*{15}{c}}

\tikzstyle{startstop} = [ellipse, rounded corners, minimum width=2cm, minimum height=1cm,text centered, draw=black, fill=gray!30]
\tikzstyle{io} = [trapezium, trapezium left angle=70, trapezium right angle=110, minimum width=2cm, minimum height=1cm, text centered, draw=black, fill=blue!30,text width=8 em]

\tikzstyle{process} = [rectangle, rounded corners, minimum width=3cm, minimum height=1cm, text centered, draw=black, fill=green!30]
\tikzstyle{process2} = [rectangle, rounded corners, minimum width=3cm, minimum height=1cm, text centered, draw=black, fill=green!30, text width=12.3em]
\tikzstyle{process4} = [rectangle, rounded corners, minimum width=3cm, minimum height=1cm, text centered, draw=black, fill=red!30, text width=20em]
\tikzstyle{process3} = [rectangle, rounded corners, minimum width=3cm, minimum height=1cm, text centered, draw=black, fill=red!30]
\tikzstyle{decision} = [diamond, minimum width=3cm, minimum height=1cm, text centered, draw=black, fill=yellow!30,text width=5em, inner sep=0pt]
\tikzstyle{arrow} = [thick,->,>=stealth]
\tikzstyle{coord}=[coordinate, node distance=6mm and 25mm]

\begin{tikzpicture}[node distance=2.2cm]

\node (start) [startstop] {\text{Start}};
\node (f1) [io,below of=start, yshift=0.6cm] {Input DATA from Main Program};
\node (f2) [process,below of=f1, yshift=0.6cm] {Create temporary matrices of density and velocity for fluid and gas };
\node (f3) [process3,below of=f2, yshift=0.6cm] {Compute density and velocity on every cell for fluid and gas};
\node (f5) [process3,below of=f3, yshift=0.6cm] {Compute the comon velosities shared among fluid and gas phases};
\node (f6) [process3,below of=f5, yshift=0.6cm] {Compute the interaction potential of each cell on both phases};
\node (f7) [process,below of=f6, yshift=0.6cm] {Neglect bubbles direct interaction on each other};
\node (f13) [decision,below of=f7, yshift=-0.4cm] {If any intersected Affected zone};
\node (f11) [process2,right of=f13, yshift=0cm,xshift=2.2cm] {Locate oxide network barrier wall in ineraction area on equal velocity cells of  each bubble interact domain};
\node (f10) [process2,below of=f11, yshift=0.2cm] {Neglect effect from other side of barrier wall};

\node (f9) [process4,below of=f13, yshift=-1cm] {Compute the final velocities due to the interaction potential and external forces};
\node (f14) [io,below of=f9, yshift=0.6cm] {Output DATA to Main Program};
\node (f15) [startstop,below of=f14, yshift=0.6cm] {Finish};

\draw [arrow] (start) -- (f1);
\draw [arrow] (f1) -- (f2);
\draw [arrow] (f2) -- (f3);
\draw [arrow] (f3) -- (f5);
\draw [arrow] (f5) -- (f6);
\draw [arrow] (f6) -- (f7);
\draw [arrow] (f13.east) --node[anchor=south, near start] {Yes} (f11);
\draw [arrow] (f13) --node[anchor=west, near start] {No} (f9);
\draw [arrow] (f7) -- (f13);
\draw [arrow] (f11) -- (f10);
\draw [arrow] (f10) |- (f9);
\draw [arrow] (f9) -- (f14);
\draw [arrow] (f14) -- (f15);

\end{tikzpicture}

  \end{tabular}
\caption{Modified Shan-Chen Algorithm \scriptsize(Green: User Codes, Red: Open Source Codes)}
\label{fig:shanchenflowchart}
\end{figure*}
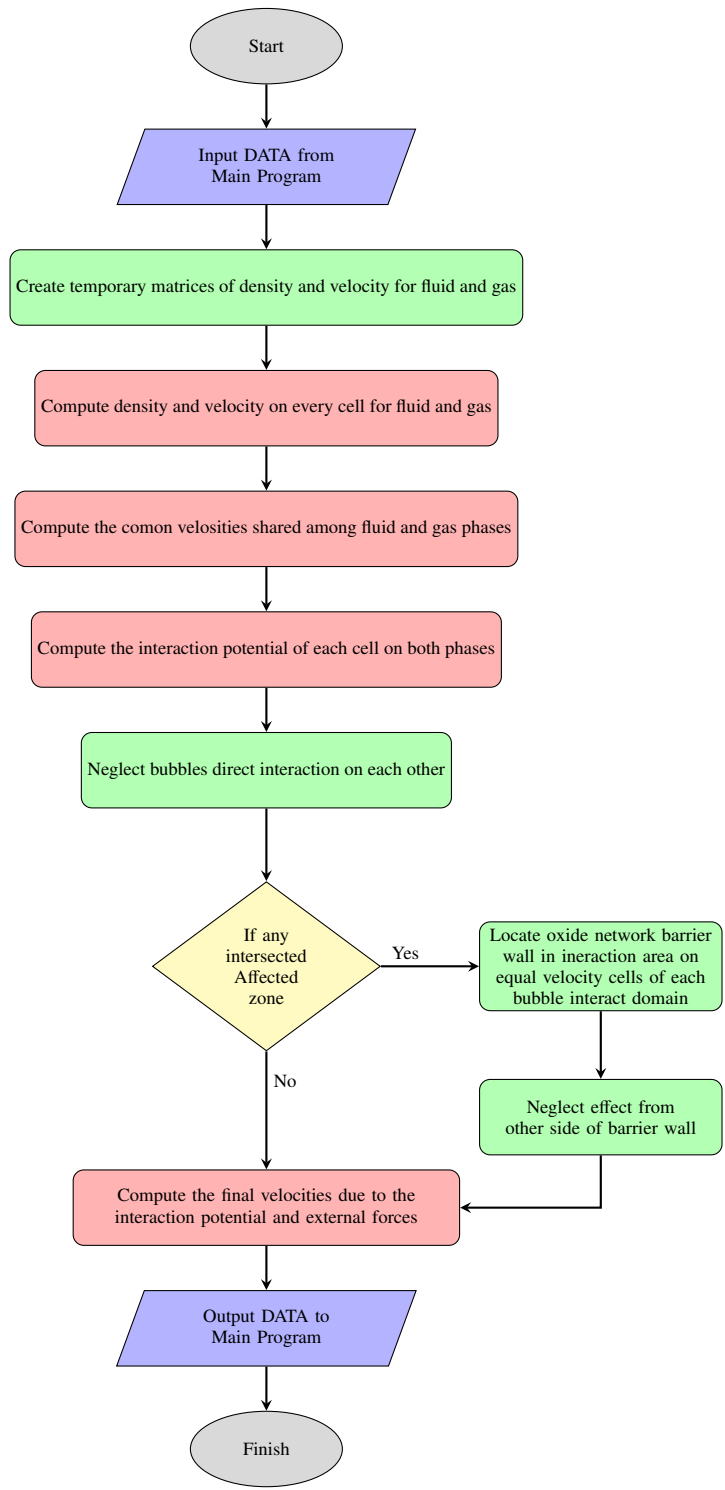

To cover this phenomenon, some simple conditions are defined. First of all, it is assumed that each bubble interacts with liquid domain only, i.e. any numerical or logical conflicts between the bubbles are neglected. Secondly, each bubble possesses an interaction zone in the liquid phase as a result of its dynamic and velocity vectors. When these domains reach each other, the attracting force between the bubbles begins its performance. As mentioned earlier, a barrier of oxide\added{-}networks is formed between these domains that prevents bubbles' coalescence. This effect could be modeled as an imaginary pressure in thin walls. By a simple condition the oxide\added{-}network effect can be simulated in the LBM code. This condition states in order to calculate corresponding Shan-Chen force at each lattice point when the interaction zones of bubbles collide, one would use the nearest bubble to that point and the effects of the rest of near bubbles are ignored, because their effect is practically neutralized by the oxide\added{-}network. This statement yields \replaced{acceptable}{reasonable}  results in the final simulation. The modified Shan-Chen algorithm is represented in Fig. \ref{fig:shanchenflowchart}. 

The computational procedure is conducted by two separate lattices, one for the melt and the other for the gas. This separation requires the utilization of the pseudopotential function, described in Eq. \ref{pseudopotential} and at the end of each time step, the lattices are coupled. Velocity and density at each lattice point is computed by Eq. \ref{eq:vel_dens}. 

Next step is the detection of the lattice points having the material between the melt and the gas (according to their velocity and density) and computing a new velocity for these points: 

\begin{equation}
\textbf{u}_{total} = \frac{\textbf{u}_{melt}+\textbf{u}_{gas}}{\sum \rho}
\end{equation}

Now the interaction potential could be calculated at each lattice point of the phases:

\begin{equation}
\xi = \rho c_s^2 + \frac{G}{6}\psi^2
\end{equation} 

The final stage is to compute the velocities according to the calculated interaction potential and the external forces. But in the modified model, a correction is applied on the computed values. New lattices are created for each bubble, and contribution of each lattice (i.e. each bubble) to the velocity of the desired point in liquid lattice is computed:

\begin{equation}
\xi_{gas} = \left\{
\begin{array}{l l} 0& \qquad \sum_{k=1}^{n} |\xi_{k}|=0 \\[0.1in] Max \left( \xi_{1}, \cdots, \xi_{n}   \right) & \qquad \frac{Max \left( |\xi_{1}|, \cdots, |\xi_{n}|   \right)}{Max \left( \xi_{1}, \cdots, \xi_{n}   \right)} = 1 \\[0.1in] Min \left( \xi_{1}, \cdots, \xi_{n}   \right)  & \qquad \frac{Max \left( |\xi_{1}|, \cdots, |\xi_{n}|   \right)}{Max \left( \xi_{1}, \cdots, \xi_{n}   \right)} \neq 1    \end{array}
\right.
\end{equation}
where $ n $ is the number of bubbles in the domain. And the final velocities could be calculated:

\begin{equation}
\textbf{u}_{gas} = \textbf{u}_{total} + \tau_{gas} \left( F_{gas} - G \xi_{melt}   \right)
\end{equation}

\begin{equation}
\textbf{u}_{melt} = \textbf{u}_{total} + \tau_{melt} \left( F_{melt} - G \xi_{gas}   \right)
\end{equation}

When the interaction condition is applied to simulate the oxide network, the bubbles would never cross the oxide network barrier and no coalescence occurs in the main domain. To solve this problem, another condition should be \replaced{defined}{define}. If this condition is defined based on experimental data, it would dictate\deleted{s} if the distance between two bubbles reach a critical number, the oxide network wall condition would be removed, which leads the normal procedure of bubble dynamics to proceed. In mathematical and computational declarations, various variables could be processed in each time step in order to determine when to remove the wall condition. In this study, the pressure field and its second derivative in normal direction are the chosen criteria to develop a condition to model removal of thin film and bubble coalescence. This condition is removing the wall when the second derivative of pressure equals zero:

\begin{equation}
\eta_m = \left\{
\begin{array}{l l}
1& \quad \frac{d^2 p}{dn^2} \neq 0 \\[0.1in]
0& \quad \frac{d^2 p}{dn^2} = 0
\end{array}
\right.
\end{equation}

 By this removal of interaction condition, bubbles rapidly merge and their dynamic effect on liquid domain could be observed. 

\section{Experimental model}
\label{experimental}
\subsection{Foam formation dynamics}
\label{foam-dynamics}

Metal foams consist of bubbles solidified just before the coalescence stage\deleted{; consequently the final structure of the foams is similar to the semi-stable structure before the solidification. The behavior of gas bubbles is not identical in different liquids and depends primarily on the active mechanisms in the continuous phase}.

Bubble coalescence is an important step in foam formation process, in which the bubbles are merged by two different mechanisms. The first mechanism is the diffusion of gas from small bubbles to the bigger ones, known as Oswald Ripening, and is more understandable. The second mechanism is thin wall rupture. \deleted{As the bubbles in the continuous phase approach each other,} \replaced{A}{a} thin film is formed between \added{the bubbles while approaching each other}\deleted{them}. The characteristic of this thin film is the same as the continuous phase; for example\added{,} in aqueous solutions, the interaction of surfactants in bubble surface is the reason of the existence of this film \cite{RefEmul}.

The behavior of bubbles depends on the low-range and high-range forces which are the result of the type of surfactant, temperature and other components in the solution. In aqueous solutions, this mechanism depends primarily on the type of surfactant and \replaced{it has little dependence on}{is less dependent to} the characteristic of the phases. \replaced{However,}{But} in some cases, such as air bubbles in oil, where the van-der Waals forces have low contribution in the interaction, the mechanism is different. In these solutions, the surfactants would be replaced by stabilizers. Presence of suspended impurity particles could result in a delay in coalescence stage \cite{RefEmul}.

Thin film rupture phenomena would be described in two different mechanisms. One of these mechanisms is the nucleation of a void and its growth due to surface tension forces. In this theory, in micro scale, a hole is formed randomly in the thin film. This formation consumes the energy $ E_r $ ($ 2 \pi \Gamma r - 2 \pi \gamma r^2 $). If the size exceeds the critical radius $ r^* $ ($ \frac{\gamma}{2\Gamma} $), then stability is expected for the formed void, but in sizes less than the critical value, the void would be removed. Thus this mechanism requires activation energy $ E_a=E(r^*) $ to start, i.e. the nucleation of void is a thermal activated mechanism and would be described by Arrhenius equation \cite{RefEmul}.

The second mechanism, which is one of the considerations in present study, considers an instability similar to Spinodal decomposition. In this mechanism, if the thickness of the thin film falls below a critical value due to drainage, a perturbation will occur. Now the instability in the thin film is appeared as the wavelength of this vibration, exceeding the critical wavelength, which leads to the rupture of the film. This critical wavelength would be calculated as below:

\begin{equation}
\lambda_c = \sqrt{\frac{-2 \pi^2 \gamma}{d^2V/dh^2}  }
\end{equation}
where $ \gamma$ is the surface energy and $V(h)$ the free energy of interface as a function of the thickness. If one considers the thin film as a cylinder with radius $R_f$ and thickness $h$ which is enclosed by two interface with surface tension $\gamma$, then the critical thickness of drainage would be:

\begin{equation}
h_c = 0.22 \sqrt[4]{\frac{AR_f^2}{f\gamma}}
\end{equation}

As the thin film thickness reaches this value, the rupture will occur. The time taken from the instability to the rupture could be calculated too: 

\begin{equation}
\tau = \frac{96 \pi^2 \gamma \eta h_c^5}{A^2}
\end{equation}

If the limitation could be neglected, the thickening would continue up to the formation of a molecular wall. But in the presence of stabilizer particles, surfactants or oxide films, there would be a different situation. The described mechanism is more suitable for rapid growth, in which the system has no surfactant. Consequently, this mechanism is appropriate to be used in present investigation. 

In metal melts, the oxide films and stabilizer particles are presented, which means it's more probable for the second mechanism to occur. Thus one of the objectives of developed code is to model the instability in the thin wall formed between the bubbles.

\subsection{Foam production}

To study the structure of a porous metal foam, Formgrip method \replaced{i}{I}s chosen to produce aluminum metal foam. This method is a combination of powder metallurgy and melting method. At first, precursor of \replaced{$A356$}{$Al356$}with $TiH_2$ is produced by melting. Then the precursor containing bubble nuclei \replaced{is}{was} placed in a mold and \replaced{is}{was} heated in furnace. Near \replaced{the}{its} liquidus temperature\added{,} the  precursor suddenly begins to blow\deleted{ because of bubble growth}. Bubble growing continues until they reach each other and before \deleted{the} coarsening stage, the part is solidified. \added{Then it is prepared for metallography images from cross-section of $A356$ foam after etching process.} \replaced{Finally,}{By cutting and comparing} the \replaced{experimental metallography images}{metallographic structure} \deleted{of the produced aluminum metal foam} \added{will be compared with bubbles images obtained from a} \deleted{with the results of the} simulation code \added{that is} developed in the present study\replaced{.}{,} \replaced{Also accuracy of the present code is}{the accuracy of the code can be} evaluated \added{as quantitative}.

For producing the precursor, around 350g aluminum 356 alloy is melted in the furnace, around 1.5 \%wt. blowing agent is mixed with aluminum powder with fraction of 0.5. For improving blowing agent wetting property in aluminum alloy melt and better uniform distribution, at 700 $^\circ C$ the blowing agent mixture is added to the melt. In the next step, the furnace temperature is set to 600 $^\circ C$. When the temperature reaches 620 $^\circ C$, the mixer rpm is set to 1500 and stirs for 1-2 min. In this step, more stirring causes more gas release. The resulting melt is rapidly casted into a metal mold.

\begin{figure*}
        \centering
                \includegraphics[width=5in]{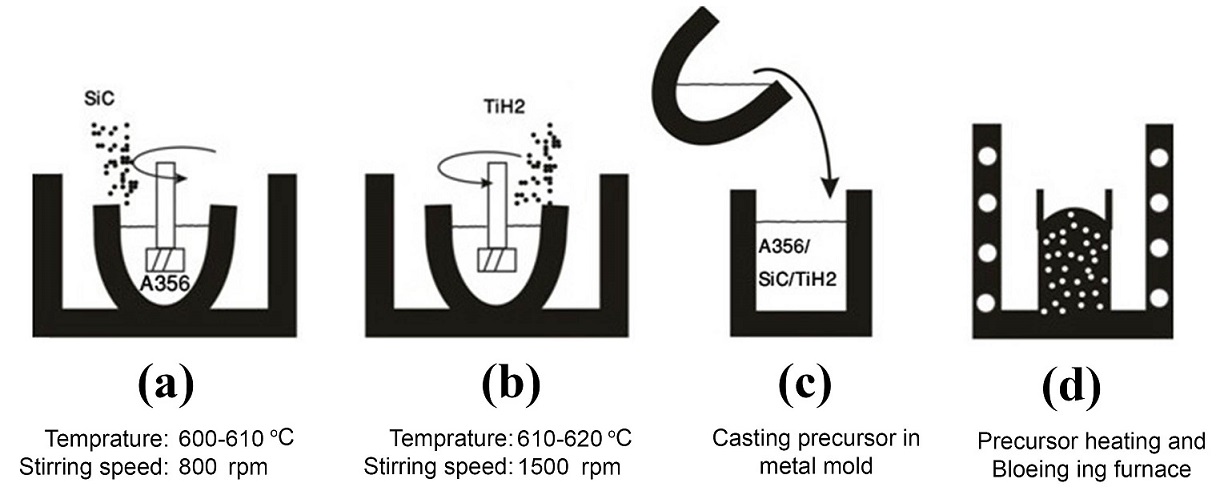}
        \caption{Schematic of foam producing steps in formgrip method}
        \label{fig:4}
\end{figure*}

The produced solid is called precursor. The precursor is cut for \replaced{foaming process}{sampling} according to the mold size (a cylinder with a radius of 40mm and 80mm height). Then the mold and the precursor are heated in furnace to blow. In the present study, the effect of blowing temperature on the stability of foam for 675, 725 and 775 $^\circ C$ is investigated. For each temperature, 2-3 samples are produced to estimate the optimized foam processing time for producing foam with minimum density and stable cell structure. Schematic of foam producing steps is shown in Fig. \ref{fig:4}.

\subsection{Experimental test and validation}
\label{metallo}

In order to determine the accuracy of the simulation results, simulated cellular structure of the present code have been compared with the experimental results. Thus, following the production of $A356$ foam, samples were sliced and mounted by black epoxy resin. For this purpose\added{,} the samples were washed with alcohol, then heated up by a dryer for better resin penetration in the foam cells, and are finally mounted. After curing, the samples' surfaces are polished with 230 to 800 grit sandpapers. Samples' surfaces need to be polished in order to give a clear picture of cellular structures. Images of samples cross sections are taken by SONY digital camera with 300dpi resolution.

\section{Results}
\label{results}
\begin{figure}[htbp]
\centerline{\includegraphics[width=2in]{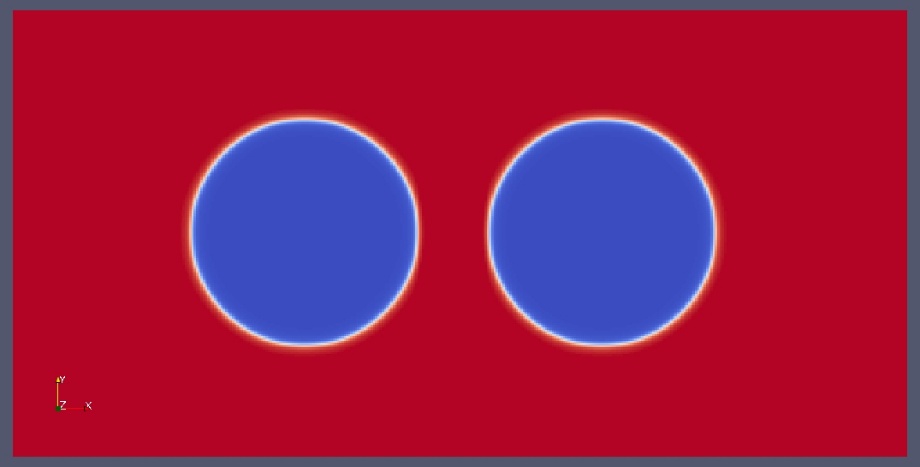}}
\caption{Initial state of the validation simulation, two gas bubble in aluminum melt}
\label{fig:initial}
\end{figure}

\begin{figure}
\centerline{\includegraphics[width=3in]{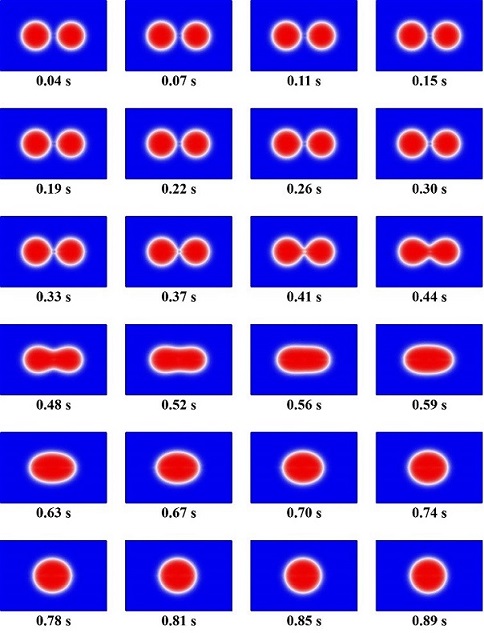}}
\caption{Simulation of two inline bubbles by finite element method and level set model in COMSOL package}
\label{fig:5}
\end{figure}

\begin{figure}
\centerline{\includegraphics[width=3in]{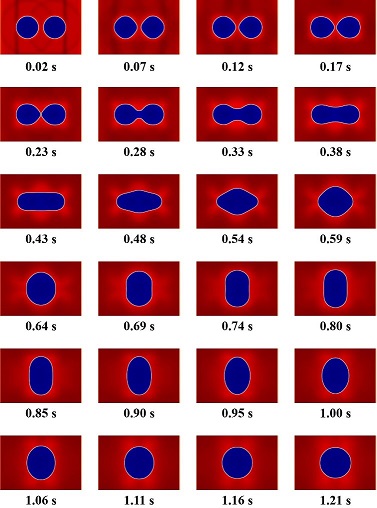}}
\caption{Simulation of two inline bubbles by LB method and Shan-Chen multiphase model in OpenLB open source code}
\label{fig:6}
\end{figure}

The coarsening of two in line bubbles \added{shown in Fig. \ref{fig:initial}} without gravity have been simulated by two different methods. Fig. \ref{fig:5} shows the 	results of the interaction between two bubbles simulated by COMSOL \replaced{commercial}{comerical} software using finite element method and level\added{-}set model. Results of LB method with OpenLB open source code is shown in Fig. \ref{fig:6}. Multiphase model is Shan-Chen and lattice is D2Q9. Simulation conditions can be seen in Table.\ref{tab:1}.

\begin{table*}[htbp]
\center \small \caption{Simulation condition for two in line bubbles}
\begin{tabular}{|c|c|c|c|}
\cline{3-4}  \multicolumn{2}{c|}{}& COMSOL & OpenLB \\ 
\hline  \multicolumn{2}{|c|}{Domain dimension (mm)}  & $ 20 \times 30 $  & $ 20 \times 30 $  \\
\hline  \multicolumn{2}{|c|}{Fluid density ($ \frac{g}{cm^3} $)} & 2.7 & 2.7 \\ 
\hline  \multicolumn{2}{|c|}{Gas density ($ \frac{g}{cm^3} $)} & 0.089 & 0.089 \\ 
\hline  \multirow {6}{*}{Initial condition}  & \multirow {3}{*}{Melt} & $ V=0 \: mm/s $ & $ V=0 \: mm/s $ \\
&& $ P=1 \: atm $& $ P=1 \: atm $\\
&& $ \mu = 1.10 \times 10^{-3} \: N.s/m^2 $& $ \mu = 1.10 \times 10^{-3} \: N.s/m^2 $\\ \cline{2-4}
& \multirow {3}{*}{Bubble} & $ V=3 \: mm/s $ & $ V=3 \: mm/s $  \\
&& $ P=1 \: atm $& $ P=1 \: atm $\\
&&$ \mu = 1.87 \times 10^{-5} \: N.s/m^2 $&$ \mu = 1.87 \times 10^{-5} \: N.s/m^2 $\\
\hline  \multicolumn{2}{|c|}{Boundary condition} & Slip & Mirror \\ 
\hline  \multicolumn{2}{|c|}{Initial bubbles D (mm)}  & 8 & 8 \\ 
\hline  \multicolumn{2}{|c|}{Final bubbles D (mm)}  & 11.29 & 11.30 \\ 
\hline 
\end{tabular} 
\label{tab:1}
\end{table*} 


\begin{figure*}
\centerline{\includegraphics[width=5in]{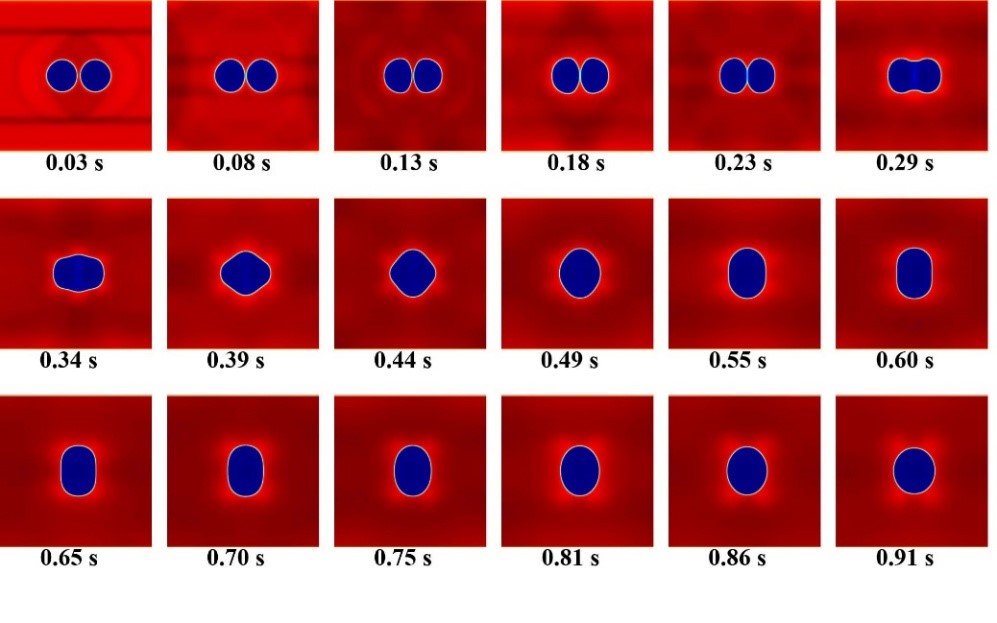}}
\caption{Coarsening simulation of two in line bubbles by LB method and modified Shan-Chen multiphase model in OpenLB}
\label{fig:7}
\end{figure*}

\begin{figure}
\centerline{\includegraphics[width=3.5in]{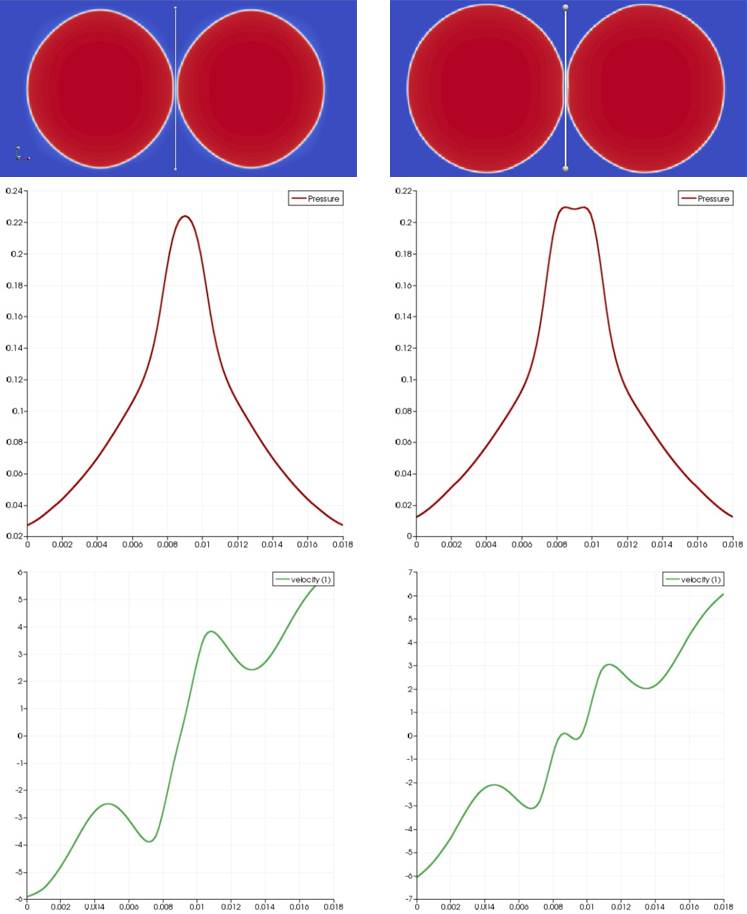}}
\caption{Plot of pressure and velocity across a vertical line in the middle of two bubbles. Left: before instability, Right: after instability. Red line indicates the pressure and green line is vertical velocity.}
\label{fig:plot}
\end{figure}

In present study, modified Shan-Chen model in LB method is used to simulate the behavior of two in line bubbles in a foam like situation. Simulation conditions are listed in Table. \ref{tab:2} and results are shown in Fig. \ref{fig:7}. In Fig. \ref{fig:plot}, the result of the pressure and velocity fields is demonstrated by the plot of the values across a vertical line, before and after the instability caused by the \replaced{perturbation}{purturbation} field of the interacting bubbles.

\begin{table}
\center \small \caption{Simulation condition for two inline bubbles by LB method and modified Shan-Chen model}
\begin{tabular}{|c|c|c|}
\cline{3-3}  \multicolumn{2}{c|}{}& OpenLB \\ 
\hline  \multicolumn{2}{|c|}{Domain dimension (mm)}  & $ 45 \times 45 $  \\
\hline  \multicolumn{2}{|c|}{Fluid density ($ \frac{g}{cm^3} $)} & 2.7 \\ 
\hline  \multicolumn{2}{|c|}{Gas density ($ \frac{g}{cm^3} $)} & 0.089\\ 
\hline  \multirow {6}{*}{Initial condition}  & \multirow {3}{*}{Melt} & $ V=0 \: mm/s $ \\
&& $ P=1 \: atm $\\
&& $ \mu = 1.10 \times 10^{-3} \: N.s/m^2 $\\ \cline{2-3}
& \multirow {3}{*}{Bubble} & $ V=3 \: mm/s $\\
&& $ P=1 \: atm $\\
&&$ \mu = 1.87 \times 10^{-5} \: N.s/m^2 $\\
\hline  \multicolumn{2}{|c|}{Boundary condition} & Mirror \\ 
\hline  \multicolumn{2}{|c|}{Initial bubbles D (mm)} & 8 \\ 
\hline  \multicolumn{2}{|c|}{Final bubbles D (mm)}  & 11.30 \\ 
\hline 
\end{tabular} 
\label{tab:2}
\end{table} 

\begin{figure}
\centerline{\includegraphics[width=3in]{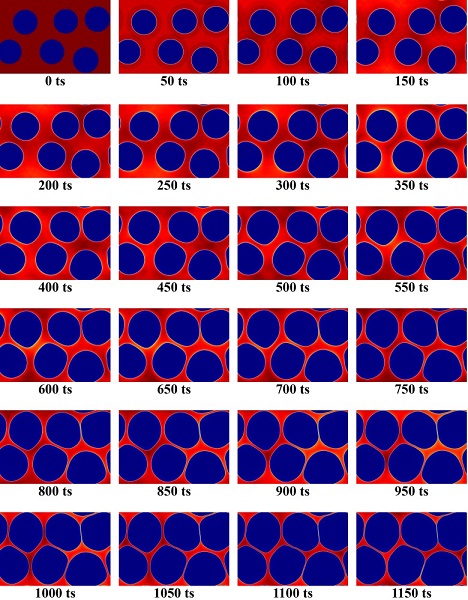}}
\caption{Simulation results of foaming stage for small domain of metal foam by LB method and modified Shan-Chen model with mirror boundary condition in OpenLB}
\label{fig:8}
\end{figure}

\begin{figure}
\centerline{\includegraphics[width=2.5in]{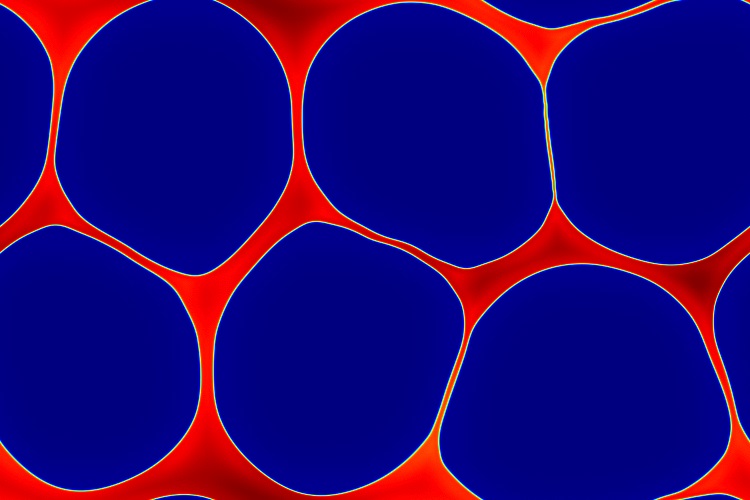}}
\caption{Bubble cell walls thinning simulation during growth step of a small foam domain (top left bubbles)}
\label{fig:9}
\end{figure}

\begin{figure}
\centerline{\includegraphics[width=2.5in]{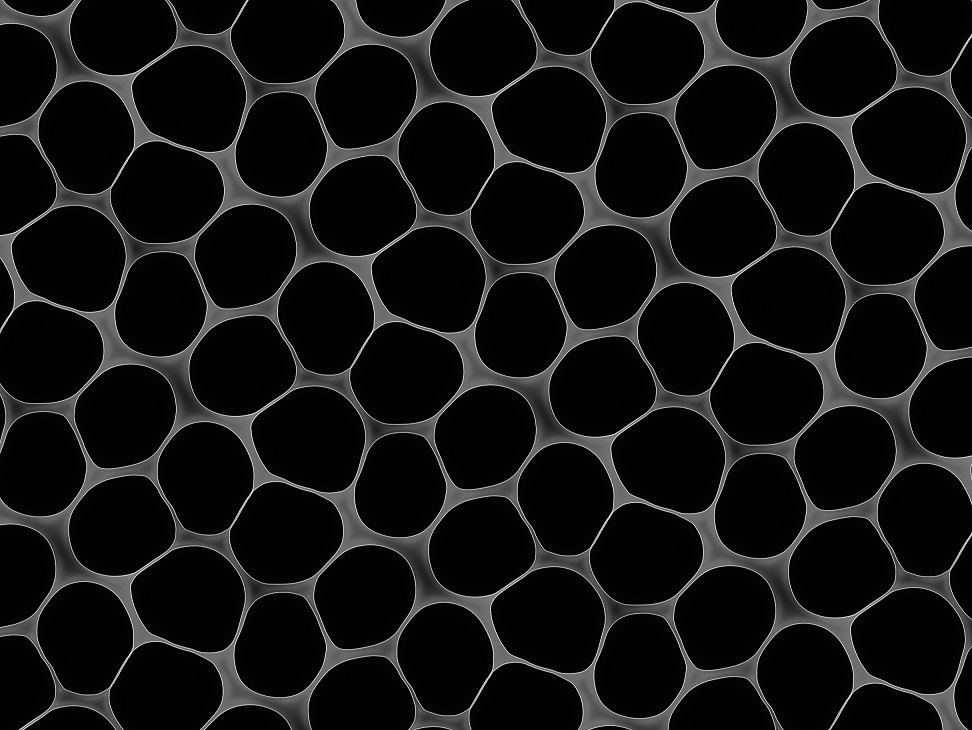}}
\caption{Imaginary metallographic structure of porous aluminum metal foam that created from mirror repeating of Fig. \ref{fig:9}}
\label{fig:10}
\end{figure}

\begin{figure*}
        \centering
                \includegraphics[width=4in]{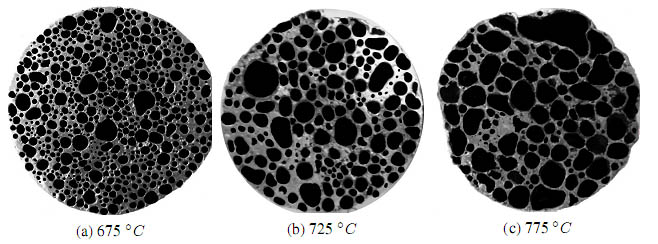}
        \caption{Metallographic image of aluminum A356 metal foam}
        \label{fig:11}
\end{figure*}

\begin{figure*}
\centerline{\includegraphics[width=4in]{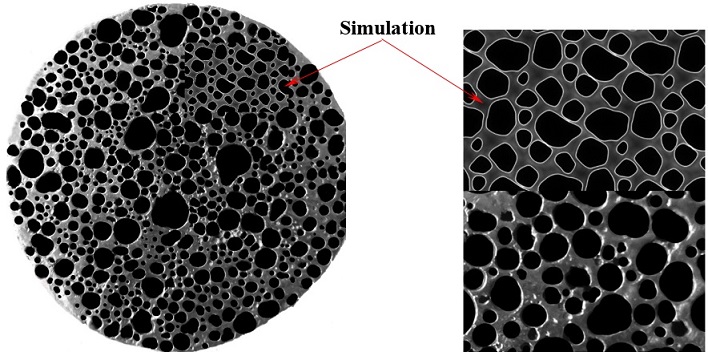}}
\caption{Experimental and simulation metallographic structure of aluminum metal foam produced by formgrip in 675 $^\circ C$}
\label{fig:12}
\end{figure*}

\begin{figure*}
\centerline{\includegraphics[width=3.5in]{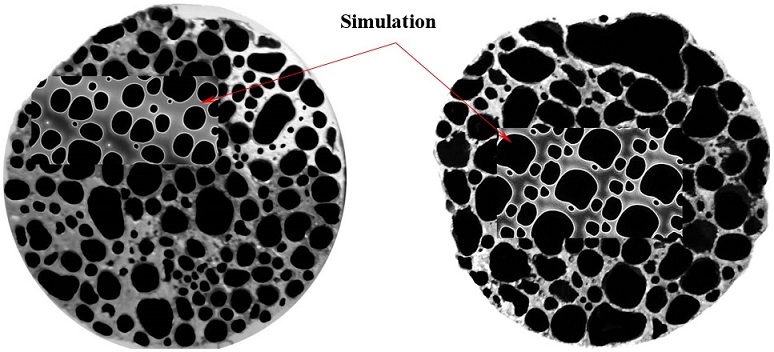}}
\caption{Experimental and simulation metallographic structure of aluminum A356 foam  (left: 725 $^\circ C$ and right: 775 $^\circ C$)}
\label{fig:13}
\end{figure*}

The Simulation results of bubble	 growth in a small domain of closed-cell aluminum foam structure with 6 primary bubbles in the mirror boundary conditions are shown in Fig. \ref{fig:8} and the simulation data are listed in Table. \ref{tab:3}. In Fig. \ref{fig:9} magnified picture of aluminum foam cellular structure is shown to evaluate the accuracy of this study's  code in regards to the detection of cell wall thinning stage in bubble coarsening (see top right of Fig. \ref{fig:9}). Small domain of Fig. \ref{fig:9}, as mentioned before is simulated by mirror boundary condition which dictates that if any melt comes out of a wall it has to come from an opposite one. Thus, by repeating the simulated mirror domain, domain of a few millimeters can be transformed, by a good approximation, to a several centimeters domain. The result of such repetitions is shown in Fig. \ref{fig:10}. Blue and red color of phase contours have been changed to gray scale to resemble the color of aluminum foam cross sections. In this case, it could be assumed that the speed of solidification is fast enough to freeze the molten aluminum foam into solid state as the cell structure maintains its molten state. Thus Fig. \ref{fig:10} can be a part of a solid aluminum foam structure. In \replaced{other}{simpler}	 words, Fig. \ref{fig:10} can be considered as \replaced{a simulation metallography}{an imaginary metallographic} picture of porous aluminum foam structure. Fig. \ref{fig:11} \replaced{shoes the}{is a} experimental	 metallograph\replaced{y}{ic} pictures of aluminum $A356$ foam \added{that} produced by Formgrip method in 675, 725 and 775 $^\circ C$,  \deleted{taken} for comparison with \replaced{the}{imaginary} simulated metallographic pictures. The results of this comparison are shown in Fig. \ref{fig:12} and \ref{fig:13}.

\begin{table}
\center \small \caption{Simulation conditions of small domain of metal foam by LB method and modified \replaced{Shan-Chen}{shan-chen} model}
\begin{tabular}{|c|c|c|}
\cline{3-3}  \multicolumn{2}{c|}{}& OpenLB \\ 
\hline  \multicolumn{2}{|c|}{Domain dimension (lattice parameter)}  & $ 750 \times 500 $  \\
\hline  \multicolumn{2}{|c|}{Fluid density ($ \frac{g}{cm^3} $)} & 2.7 \\ 
\hline  \multicolumn{2}{|c|}{Gas density ($ \frac{g}{cm^3} $)} & 0.089\\ 
\hline  \multirow {6}{*}{Initial condition}  & \multirow {3}{*}{Melt} &  $ \mu_{675} = 1.20 \times 10^{-3} \: N.s/m^2 $ \\
&&  $ \mu_{725} = 1.10 \times 10^{-3} \: N.s/m^2 $\\
&& $ \mu_{775} = 1.02 \times 10^{-3} \: N.s/m^2 $\\ \cline{2-3}
& \multirow {3}{*}{Bubble} &  $ \mu_{675} = 1.82 \times 10^{-5} \: N.s/m^2 $\\
&&  $ \mu_{725} = 1.87 \times 10^{-5} \: N.s/m^2 $\\
&&$ \mu_{775} = 1.93 \times 10^{-5} \: N.s/m^2 $\\
\hline  \multicolumn{2}{|c|}{Boundary condition} & Mirror \\ 
\hline 
\end{tabular} 
\label{tab:3}
\end{table}

\section{Discussion}
\label{dis}
In Lattice Boltzmann method, the macroscopic properties of the domain of interest could be predicted by solving the LB equations at mesoscopic scale. In this investigation, this numerical method is used and the Shan-Chen scheme for multiphase modeling is modified to present a code to predict the behavior of metal foams. In available commercial CFD codes, the modeled behavior of bubbles is not similar to the dynamics seen in experimental observation\deleted{, as discussed in \replaced{page}{section}  \pageref{interaction}}. By modifying the Shan-Chen model (which is one the most accurate models in multiphase LB simulations), improved results in field of bubble dynamics would be achieved.

In order to validate and compare the developed code, two bubbles at growth stage in a container are considered and the dynamics of interaction is determined \added{by} using a conventional CFD code, \deleted{the} unmodified LB code and present code. The boundaries of the domain are assumed to be closed for the gas phase and opened for the liquid phase, which causes the extra liquid exit the domain as the gas bubbles grow. The interaction begins as the volume of the bubbles increases. The initial state of this model is represented in Fig. \ref{fig:initial}.

\replaced{According to Fig. \ref{fig:5} and Fig. \ref{fig:6},}{Comparison of the} \added{obtained simulation} results \replaced{from}{of} the LBM method with unmodified Shan-Chen model \replaced{by}{in} OpenLB \added{code,} and \added{the} \replaced{FEM}{finite element method} with level set model \replaced{by}{in} COMSOL software \replaced{have a same}{demonstrate the similarity in the} behavior \added{during interaction} of two bubbles\deleted{ interaction}. \deleted{This is presumable from comparison between Fig. \ref{fig:5} and Fig. \ref{fig:6}. }In both methods, as the bubbles collide, a tip is formed on bubbles boundary \added{at the interface zone between two bubbles}. This shows the absorption force of bubbles  that moves bubbles toward each other for merging and decreasing surface energy. \added{Obtained quantitative results  from both simulations} \deleted{Result comparison is} listed in Table. \ref{tab:4}.
 
 \begin{table}[htbp]
\center \small \caption{Simulation result comparison for two in line bubbles}
\begin{tabular}{|c|c|c|c|c|}
\cline{3-5}  \multicolumn{2}{c|}{}& COMSOL & OpenLB & Modified code \\ 

\hline  \multicolumn{2}{|c|}{Boundary condition} & Slip & Mirror & Mirror\\ 
\hline  \multicolumn{2}{|c|}{Initial bubbles D (mm)}  & 8 & 8 & 8 \\ 
\hline  \multicolumn{2}{|c|}{Final bubbles D (mm)}  & 11.29 & 11.30 & 11.30 \\ 
\hline  \multicolumn{2}{|c|}{Merging time (s)}  & 0.26 & 0.25 & 0.31 \\ 
\hline 
\end{tabular} 
\label{tab:4}
\end{table} 
 
\replaced{These}{Thses} results are \replaced{factual}{true} for every liquid without any solid impurity. But as mentioned before this behavior is not acceptable for metal\added{s} melt, especially in aluminum and aluminum alloy\added{s} foaming process. In metals there is no tip at the intersection of two bubbles. Therefore, for bubbles growth in metal melt, a new model should be developed. In this study, this objective is accomplished by using a modified version of the Shan-Chen model in LBM simulation of metal foaming process.  The result of the simulation performed using this new code for aluminum melt compared with COMSOL and OpenLB software is shown in Table. \ref{tab:4}. 
The result shown in Fig. \ref{fig:6} means that the bubble dynamics equations are well satisfied and the result domain shows behavior based on these equations. 
But the behavior of bubbles in the interaction is not the one expected from metal melts due to the presence of impurities \added{and the particles-network} as discussed before. In Fig. \ref{fig:7}, \replaced{simulation}{the} result\added{s} \added{illustrated based on} \deleted{of simulation performed with} the modified Shan-Chen model\deleted{ is represented}. As \replaced{shown,}{seen} in \replaced{this}{the} figure, \replaced{at interface of}{in} the bubble interaction\deleted{ interface}, a thin film is formed and \added{the} merging \added{phenomenon} does not occur until a specific criterion is met.\deleted{ This behavior is the expected dynamics of bubble interaction in metal melts.}

Furthermore, to complete the validation results of the developed code, the pressure and velocity values across a vertical line is shown in Fig. \ref{fig:plot}. \deleted{As the film in the middle of the bubbles thickens,} \replaced{It is}{it's} observed that \replaced{for}{in} a specific thickness \added{an instability will} \deleted{(which is loosely dependent on the size of the bubbles), an instability is }occur, \added{which is dependent on the size of the bubbles.} \deleted{as discussed in \replaced{page}{section} \pageref{foam-dynamics}.} This instability could be seen in the pressure profile \replaced{(right image of Fig. \ref{fig:plot})}{and is similar to the Spinodal decomposition}. Thus, in the beginning of the instability, the second derivative of pressure \added{($ \nabla^2P $)} profile in the lattice point would be zero, which indicates the initialization of \added{bubbles} wall rupture. \replaced{Then it is}{So it’s} possible and more comfortable to determine the wall rupture \added{by} using the pressure profile and \replaced{pressure}{processing its} second derivative instead of detecting the critical thickness.  By applying this criterion to the domain of foam formation, the merge condition of the gas bubbles would be detected based on the second derivative of pressure field at each lattice points, which leads to an improved modeling of metal foam formation process.

Designing an experiment that could show\deleted{s} the dynamics of two moving bubbles in aluminum melt to validate present simulation is very expensive and requires very especial equipment for casting and would be impossible due to non-transparent molten metals. \deleted{Gathering all simulation requirements in any experiment is difficult but in this specific case that has many more parameters, involves more difficulty. So to compare results of virtual world to real world of molten metal, the metal foam structure should be used. The metallographic images of cross cutting solid metal foam samples compared with the simulation results of the foam as virtual metallographic images, such as those presented in Fig. \ref{fig:10}, thus the accuracy of the simulation results would be evaluated visually. For this purpose,} \added{In Fig. \ref{fig:10} shows simulation results of a cross-section of A356 foam based on} the modified Shan-Chen method \added{at present code. In this image,} \deleted{is used to simulate} a small portion of the domain \deleted{and} \added{has been repeated by} using \replaced{periodic}{mirror} boundary condition\added{.} \deleted{to repeat these portions all together.}

\begin{figure*}
\centerline{\includegraphics[width=3.5in]{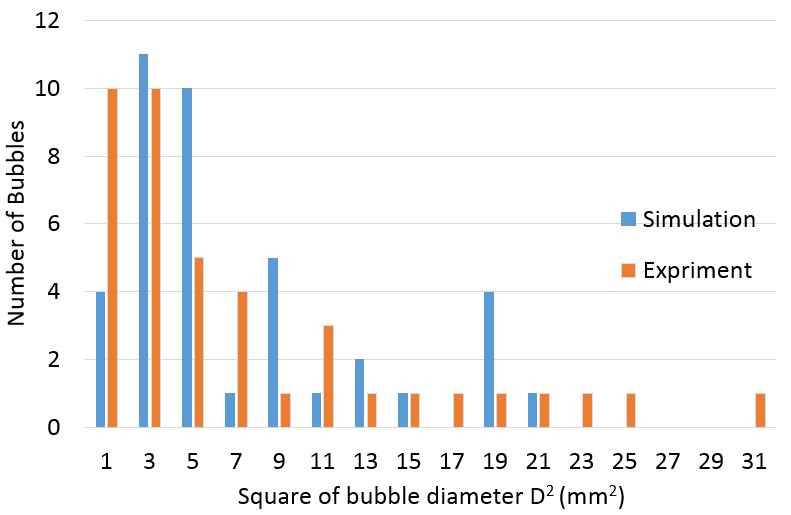}}
\caption{Bubble size distribution of simulation and experimental samples of Fig. \ref{fig:12}}
\label{fig:14}
\end{figure*}

 \begin{table}
\center \small \caption{Comparison of extracted data from simulation and experimental  metallographic cellular structures of Fig. \ref{fig:12}}
\begin{tabular}{|c|c|c|c|c|}
\cline{3-5}  \multicolumn{2}{c|}{}& simulation & experimental & \added{Error}\\ 

\hline  \multicolumn{2}{|c|}{Bubble percentage} & 44.3\% & 45.5\% & \added{2.64\%}\\ 
\hline  \multicolumn{2}{|c|}{Foam Density \tiny ($ \frac{g}{cm^3} $)}  & 1.50 & 1.47 & \added{2.04\%} \\ 
\hline  \multicolumn{2}{|c|}{Mean bubble size (mm)}  & 3.03 & 3.05 & \added{0.66\%} \\ 
\hline 
\end{tabular} 
\label{tab:5}
\end{table} 

Fig. \ref{fig:8} shows foaming stage for small domain of metal foam by time. Bubbles with random distribution began to grow after nucleation. Bigger bubbles have a \replaced{higher}{bigger} growth rate than the \replaced{others}{smallers}. Each bubble has an affected zone\added{, and}	 \replaced{w}{W}hen \replaced{these}{this} zones reach each other, bubbles interaction will begin. Simulation \added{of bubbles' growth} will continue until first cell reach\added{es the} wall rupture \replaced{criterion}{condition,} \added{as like Fig. \ref{fig:9}}. \replaced{After this time (best foaming time),}{If we decided to not to stop the simulation at this stage, due to small domain size and small deference between bubbles size,} foaming \added{simulation} process would suddenly enter the aging step and bubbles coalescence\deleted{ happen rapidly}, \added{which} leads to a \replaced{sever}{sevier} drainage in metal foam structure. \deleted{The final picture resulted from simulation of small metal foam domain with modified Shan-Chen model (Fig. \ref{fig:9}) could be repeated together due to the usage of mirror boundary condition up to the cross section of sliced real samples (like Fig. \ref{fig:10}). Resulted picture from mirror boundary condition simulation is the} \added{These} simulation metallograph\replaced{y}{ic} image\added{s could be used for predicting of experimental metallography images (as like Fig. \ref{fig:11}) and optimum foaming time}\deleted{ that can be compared with the real one (Fig. \ref{fig:11})}. These comparisons \deleted{is} are represented in Fig. \ref{fig:12} \added{for 675 $^\circ C$} and Fig. \ref{fig:13}  \added{for 725 and 775 $^\circ C$, respectively}. \deleted{Thermal condition for hydrogen gas release is set to  675 $^\circ C$ in setup parameters of simulation program. Virtual metallographic image of the simulation conducted in this temperature is compared with real one in Fig. \ref{fig:12}. Furthermore similar results for two other temperatures are shown in Fig. \ref{fig:13}.} In addition to visual results, \replaced{quantitative}{quantified} results \added{are extracted from Fig. \ref{fig:12}} \replaced{simulation data}{are also obtained from the simulations} in order to validate\deleted{ the developed numerical model}. These extracted results \replaced{are processed}{from Fig. \ref{fig:12}} \added{by using MATLAB image process\replaced{or}{ing}} \replaced{and are illustrated}{are compared} in Fig. \ref{fig:14} and Table. \ref{tab:5}. \replaced{These}{This} data show\deleted{s} a minor error between the simulation and the experiment \deleted{and} results\added{.} \deleted{of the predicted virtual metallographic cellular structure are in a acceptable agreement with real samples structure. This means}  \added{Therefore,} \replaced{present}{the developed} code \replaced{based on}{with} modified Shan-Chen model \replaced{and LB}{to solve multiphase fluid dynamics equations by lattice Boltzmann} method could simulate and predict \replaced{foamy}{cellular} structure\added{s} \replaced{for}{of} aluminum $A356$ \replaced{foams during isothermal}{resulted from} foaming process\deleted{ as well as real formgrip produced samples}.

\section{Conclusions}
\label{conc}
In this investigation, the Lattice Boltzmann Method (LBM) was utilized for understanding of foaming process by simulation of different stages of the Aluminum A356 foam production process at micro and meso scales. Therefore, to predict the structure of metal foam during foaming process, a model is established to simulate the dynamics of bubbles interaction based on development of Shan-Chen model. The presented model can consider the effect of the attraction-repulsion barriers among bubbles into the A356 aluminum foam liquid due to the solid particles network. In order to validate the presented model, results were compared by both some other reference codes and experimental data. Comparison of cellular structure obtained from the experimental route (experimental metallography) and the numerical code (simulation metallography) shows a good consistency. The presented code is also capable of simulating and presenting virtual metallography images for all aluminum alloys foams. Therefore, this software can be used for controlling and predicting density of foams combined with uniform distribution of bubbles at the metal foams.

\clearpage

\bibliography{REFBOOK}

\end{document}